# A Probabilistic Collocation Method Based Statistical Gate Delay Model Considering Process Variations and Multiple Input Switching[†]

Satish Kumar. Y,
Univ. of Arizona
satishy@email,arizona.edu

Jun Li,
junl98@yahoo.com

Claudio. Talarico,
Univ. of Arizona
Claudio@ece.arizona.edu

Janet Wang
Univ. of Arizona
wml@ece.arizona.edu

**Abstract**

*Since the advent of new nanotechnologies, the variability of gate delay due to process variations has become a major concern. This paper proposes a new gate delay model that includes impact from both process variations and multiple input switching. The proposed model uses orthogonal polynomial based probabilistic collocation method to construct a delay analytical equation from circuit timing performance. From the experimental results, our approach has less that 0.2% error on the mean delay of gates and less than 3% error on the standard deviation.*

## 1. Introduction

Since the advent of nanotechnology, the process geometries continue to shrink. The variability of circuit delay due to process variations has become a major concern. The ability to control critical device parameters is becoming increasingly difficult. On the other hand, traditional approaches are no longer able to predict the circuit performance accurately. For example, the corner based analysis, though was popular for die-to-die variations in the past, has found out to be too pessimistic to model within die variations. New statistical method based approaches are becoming the main stream of today's performance verification tools.

Recent research work attempts to incorporate statistical models into the Static Timing Analysis. The combination is thus referred as Statistical Static Timing Analysis (SSTA). Based on the techniques they use, the existing SSTA methods can be classified into block-based analysis approaches [1-7] and path-based analysis approaches [8-10]. These SSTA methods aim to get the arrival time or path delay associated probability distribution functions.

One of the key parts of SSTA is the gate delay model. Unlike the existing deterministic gate delay models [11], the new gate delay models have to include impact from process variation and multiple input switching. A recent paper from Agarwal, Dartu and Blaauw [12] found out that "the Multiple Input Switching has a greater impact on statistical timing than regular static timing analysis." The model they proposed is able to predict the mean and standard deviation of gate delays with less than 7% error and the average error is less than 2%.

This paper presents a new gate model by using orthogonal polynomial based probability collocation method. The proposed new model has several advantages over the other gate models: 1) by using orthogonal polynomials, the order of the gate delay analytical equation has been proved to be the lowest order theoretically [13]; 2) by applying different kind of orthogonal polynomials, our model applies to cases with both Gaussian distribution and non-Gaussian distribution dependant variables. 3) by using probability collocation method, the procedure to construct the analytical equation is greatly simplified. Comparing with general Monte Carlo approach, where several thousands of sampling points are needed to estimate the circuit performance, the probability collocation method (PCM) only requires a few sampling points; 4) by selecting collocation points using the selection criteria explained in later sections as sampling points, the PCM method avoids the randomness in choosing sampling points. Moreover, it also guarantees the overall estimation accuracy with regard to the valid range of dependant parameters; 5) the PCM method is applied to the circuit performance or response. It belongs to response surface approaches. Therefore, the PCM method does not require the system structure information. It treats the system as black box and predicts the system behavior from the system's response or performance.

Because of its advantages, the PCM method proposed in the current paper offers a gate delay model that can also be extended to block-level delay. Hence, the model is applicable to system-level designs. The experimental results show that our approach has less that 0.2% error on the mean delay of gates and less than 3% error on the standard deviation.

---

[†] The method reported by this paper is protected by patent

The rest of this paper adheres to the following format. Section 2 summarizes the flow of the proposed gate delay model. Section 3 discusses the methodology of PCM and its application to the gate delay modeling. Section 4 presents the results and Section 5 concludes this paper.

## 2. The Flow of Generating the New Gate Delay Model

In the Deep Sub Micron region (DSM), fabrication limitations introduce process variations. Important parameters like channel length, channel width, thickness of oxide, and critical length of device ($L, W_n, W_p, T_{ox}, \lambda$) are no longer deterministic quantities and should be treated as random variables. Consequently, the gate delay which depends on these parameters needs to be modeled as a random variable as well. The probability density function (PDF) of the gate delay is obtained by running simulations with a number of sample points of basic parameters ($L, W_n, W_p, T_{ox}, \lambda$). The PDF of each parameter ($L, W_n, W_p, T_{ox}, \lambda$) as well as the input signal delays are assumed to be normal distribution Our approach does not limit to normal distribution cases. A detailed explanation on other distribution cases can be found in Section 3.1. Figure 1 shows the new delay model diagram. Four major steps (Figure 2) in formulating the new gate delay model are: 1) representation of uncertain inputs; 2) functional approximation of model outputs by orthogonal polynomials; 3) estimation of approximation parameters (PCM technique); 4) calculation of statistical properties of outputs.

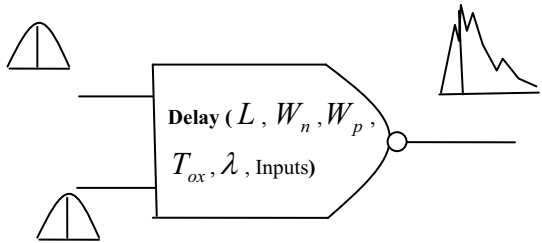

**Figure 1: New Gate Delay Model Diagram**

## 3. Orthogonal Polynomial based Probability Collocation Method

### 3.1 Representation of Uncertain Inputs

The first step in the uncertainty modeling is the representation of all the uncertain model inputs($X$) in terms of a set of "standard random variables (srv's)" $\{\xi_i\}_{i=1}^n$. $\{\xi_i\}_{i=1}^n$ is a set of independent, identically distributed (*iid*) normal random variables. '$n$' is the number of independent inputs and each $\xi_i$ has zero mean and unit variance. When the input random variables are independent, the uncertainty in the $i^{th}$ model input $X_i$, is expressed directly as a function of the ith srv, $\xi_i$ i.e., a transformation of $Xi$ to $\xi_i$ is employed. Devroye [15] presents transformation techniques and approximations for a wide variety of random variables. Table 1 presents a list of transformations for some commonly employed probability distributions. For example, if the random variable $L_{eff}$ is assumed to have a normal PDF of mean μ and standard deviation σ, then the transformation employed is μ + σ ξ, where $\xi \sim N(0,1)$.

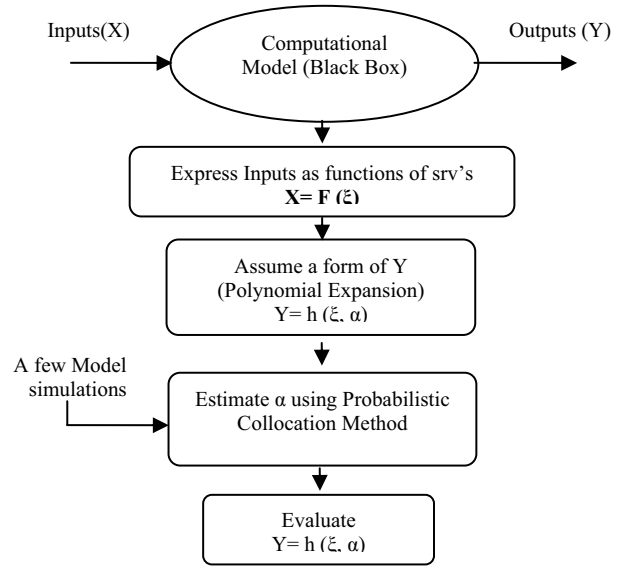

**Figure 2. General Flow of the Generation of Gate Delay Model**

**Table 1: Representation of common distributions as function of normal random variables**

| Distribution | Transformation |
|---|---|
| Uniform $(a,b)$ | $a + (b-a)\left(\frac{1}{2} + \frac{1}{2} erf\left(\xi/\sqrt{2}\right)\right)$ |
| Normal $(\mu, \sigma)$ | $\mu + \sigma \xi$ |
| Lognormal $(\mu, \sigma)$ | $\exp(\mu + \sigma \xi)$ |
| Gamma $(a,b)$ | $ab\left(\xi\sqrt{\frac{1}{9a}} + 1 - \frac{1}{9a}\right)^3$ |
| Weibull $(a)$ | $y^{\frac{1}{a}}$ |

## 3.2. Functional Approximation of Model Outputs

In this Section, the output of the model is approximated as a series expansion of Normal random variables (*srv's*), in terms of Hermite polynomials. Hermite polynomials form the best orthogonal basis if the random variables are Gaussian [14]. The series expansion is shown below,

$$y = a_0 + \sum_{i=1}^{n} a_{i1} \Gamma_1(\xi_{i1}) + \sum_{i_1=1}^{n} \sum_{i_2=1}^{i_1} a_{i1} a_{i2} \Gamma_2(\xi_{i1}, \xi_{i2}) + \sum_{i_1=1}^{n} \sum_{i_2=1}^{i_1} \sum_{i_3=1}^{i_2} a_{i1} a_{i2} a_{i3} \Gamma_3(\xi_{i1}, \xi_{i2}, \xi_{i2}) + \ldots \quad (1)$$

where $y$ is any output metric (or random output) of the model. $a_i ---'s$ are deterministic constants to be estimated and the $\Gamma_P(\xi_{i1},....,\xi_{ip})$ are multi-dimensional hermite polynomials of degree $P$ given by

$$\Gamma_P(\xi_{i1},....,\xi_{ip}) = (-1)^p e^{\frac{1}{2}\xi^T \xi} \frac{\partial^P}{\partial \xi_{i1}....\partial \xi_{ip}} e^{-\frac{1}{2}\xi^T \xi} \quad (2)$$

where $\xi$ is the vector of '$P$' iid normal random variables $\{\xi_{ik}\}_{k=1}^{p}$ that are used to represent input uncertainty.

The unknown coefficients in the Eq.(1), ($a_{ik} ---'s$) can be estimated using several techniques. For nonlinear operators with mathematically less complex equations, the unknown coefficients in the orthogonal polynomial expansion can be determined by minimizing an appropriate norm of the residual, after substituting the transformed inputs into the model equations. Gelarkin's method [14] is commonly used with weight function corresponding to the expectation of the random variables. For cases in which the model equations are not easy to manipulate, or when the model is of ``black-box'' type, the unknown coefficients can be obtained by a collocation method. This method imposes the requirement that the estimates of model outputs are exact at a set of selected collocation points, thus making the residual at those points equal to zero. The unknown coefficients are estimated by equating model outputs and the corresponding Hermite polynomial expansion, at a set of collocation points in the parameter space; the number of collocation points should be equal to the number of unknown coefficients to be found. Thus, for each output metric, a set of linear equations results with the coefficients as the unknowns, which can be solved with less computation. The following Section explains the collocation method in detail.

## 3.3. Probability Collocation Method (PCM)

In this section, we explain the concept of PCM. First, we consider the collocation method for a deterministic case and later extend to the stochastic case.

Consider a simple model with one input $x$ and one output $y$,

$$y = f(x) \quad (3)$$

where $f$ is a known explicit or implicit function. The output '$y$' can be approximated using a set of specified functions of $g_i(x)$.

$$\hat{y} = \sum_{i=0}^{N} y_i g_i(x) \quad (4)$$

where $N$ is the order of approximation. Since the approximation of $y$ may not simulate the actual model, we define the residual of the model as follows:

$$R(\{y_i\}, x) = \hat{y}(x) - y(x) \quad (5)$$

The set of coefficients in the approximation $\{y_i\}$ can be calculated by requiring that the residual and each member of $\{g_i(x)\}$ should be orthogonal to each other.

$$\int_x R(\{y_i\}, x) g_i(x) dx = 0, \quad i = 0,......,N \quad (6)$$

Equation (6) can be solved by using the Gaussian quadrature approximation.

$$\int_x R(\{y_i\}, x) g_i(x) dx \simeq \sum_{j=1}^{N} v_j R(\{y_i\}, x) g_i(x_j) \quad (7)$$

$i=0,...,N$, where $v_j$ and $x_j$ are the weights and abscissas respectively. If $v_j g_i(x_j)$ has the same sign and is not zero for all $i$ and $j$, equation (6) can be approximated by,

$$R(\{y_i\}, x_j) = 0, \quad j = 0,1,....,N \quad (8)$$

Equation (8) represents the use of the collocation method for calculating the set of coefficients $\{y_i\}$. This implies that the collocation method does not require the complete definition of the residual function. As long as we can calculate the value of the residual at several given values of inputs, we will be able to obtain the coefficients of approximation.

In stochastic models where inputs and outputs are random variables, we can extend the deterministic collocation method and can obtain an approximation of the model. The orthogonal relationship defined by equation (6) is transformed to the probabilistic space by incorporating the Joint Probability Density Function of inputs $f_{x(w)}(x(w))$:

$$\int_{x(w)} f_{x(w)}(x(w)) R(\{y_i\}, x(w)) g_i(x(w)) dx(w) = 0 \quad (9)$$

Similarly equation (8) now becomes

$$f_{x(w)}(x_j) R(\{y_i\}, x_j) = 0 \quad j = 0,....,N \quad (10)$$

If we choose $x_j$ such that $f_{x(w)}(x_j)$ is positive for all j, we can still apply equation (8) to the cases where $x$ is a random variable.

In practice, $\{g_i(x)\}$ are chosen as the orthogonal polynomials (in our case, Hermite polynomials) whose weighting function is the PDF of $x$, $f_{x(w)}(x(w))$. Therefore the collocation points $\{x_j\}$ are simply the roots of the *(n+1)* order orthogonal polynomials [13]. So, the number of collocation points available is $(d+1)^n$, where *d* is the degree of expansion and *n* is the number of inputs. The unknown coefficients of output expansion are estimated by equating model outputs (using Spice) and the corresponding Hermite polynomial expansion, at this set of collocation points $\{x_j\}$ in the parameter space. The number of sampling points (collocation Points) is equal to number of unknowns.

The approximation of output with Hermite Polynomial expansion of second degree for a two input model results in six unknown coefficients [Eq.1]. But the number of collocation points available is $(2+1)^2 = 9$. Similarly, for higher dimension systems and higher order approximations, the number of available collocation points is always greater than the number of collocation points needed, which introduces a problem of selecting the appropriate collocation points. In the next Section, we present a selection criterion for selecting the collocation points.

### 3.4. Selection Criteria for Collocation Points

In this section, we present a simple heuristic technique to select the required number of collocation points from a large number of potential candidates

The collocation points are selected so that each standard normal random variable $\xi_i$ takes the values of either zero or one of the roots of the higher order Hermite-polynomial. The zero is taken as a collation point even though if it is not a root, because the origin corresponds to the region of highest probability for a normal random variable $\xi_i$ of zero mean. For each term of the series expansion, a corresponding collocation point is selected. The collocation point corresponding to the constant is the origin, i.e., all the standard random variables ($\xi's$) are set to value zero. For terms involving only one variable, the collocation points are selected by setting all other $\xi's$ to zero value and by letting the corresponding variable take values as the roots of higher order Hermite polynomial. For terms involving two or more random variables, the values of the corresponding variables are set to the values of the roots of the higher order polynomial and so on. If more points corresponding to a set of terms are available than needed, the points which are closer to the origin are preferred as they fall in regions of higher probability. Further, when there is still an unresolved choice, the collocation points are selected such that the overall distribution of the collocation points is more symmetric with the origin. If still more points are available, the collocation point is selected randomly.

Once the coefficients used in the series expansion of the model outputs are estimated, the statistical properties of the outputs such as the density functions, moments, joint densities, join moments, correlation between two outputs or between an output and an input etc can be readily calculated. We explain in detail the calculation of statistical properties in the next Section.

### 3. 5 Statistical Properties of the Outputs

If inputs are represented as $x_i = F_i(\xi_i)$ and if outputs $y_i$ are estimated as $y_i = G_i(\xi_1, \xi_2 - - \xi_n)$, then the following steps are involved in the estimation of the statistics of the inputs and outputs.

1. Generation of a large number of samples of $[\xi_1, \xi_2, - - - \xi_{ni}]$
2. Calculation of the values of the input and output random variables from the samples.

From a set of *N* samples, the moments of the distribution of an output $y_i$ can be calculated as,

$$\eta_{y_i} = Mean(y_i) = E\{y_i\} = \frac{1}{N} \sum_{j=1}^{N} y_{i,j} \qquad (11)$$

$$\sigma_{y_i} = Var(y_i) = E\{(y_i - n_{y_i})^2\} = \frac{1}{N-1} \sum_{j=1}^{N} (y_{ij} - \eta_{y_i})^2 \qquad (12)$$

Calculation of model inputs and outputs involves evaluation of simple algebraic expressions and does not involve model runs.

### 3.6 An Inverter Delay Model by PCM

We consider an inverter to illustrate the PCM technique. Let the $L_{eff}$ and $T_{ox}$ are the two random variables and delay is the output(Y) parameter to be estimated. Let

$$L_{eff} \sim N(\mu_1, \sigma_1) \qquad T_{ox} \sim N(\mu_2, \sigma_2) \qquad (13)$$

The input random variables can be represented by standard random variables using Table 1.

$$L_{eff} = \mu_1 + \sigma_1 \xi_1 \qquad T_{ox} = \mu_2 + \sigma_2 \xi_2 \qquad (14)$$

Where $\xi_1, \xi_2$ are *iid* $N(0,1)$ random variables.

A second order Hermite Polynomial Expansion for $Y$ in terms of $\xi_1, \xi_2$ is given by

$$Y = a_0 + a_1\xi_1 + a_2\xi_2 + a_3(\xi_1^2 - 1) + a_4(\xi_2^2 - 1) + a_5(\xi_1\xi_2) \quad (15)$$

In order to estimate the 6 unknown coefficients, 6 collocation points $(\xi_{1,1}, \xi_{2,1}), (\xi_{1,2}, \xi_{2,2})$....... are selected from the roots of the 3rd degree Hermite polynomial as explained in section 3.3. These sample points correspond to the original model input samples $(Leff_1\ Tox_1) - - - (Leff_6\ Tox_6)$ as follows:

$$\begin{bmatrix} \xi_{1,i} \\ \xi_{2,i} \end{bmatrix} \rightarrow \begin{bmatrix} Leff_i \\ Tox_i \end{bmatrix} = \begin{bmatrix} \mu_1 + \sigma_1\xi_{1,i} \\ \mu_2 + \sigma_2\xi_{2,i} \end{bmatrix} \quad for\ i = 1,....,6 \quad (16)$$

After obtaining original input sample points, the model simulations using Spice are performed at the given input sample points and outputs are obtained. Then the outputs $y_1 -- y_6$ are used to calculate the coefficients $a_0 --- a_5$ by solving the following linear equation.

$$Z^T * \begin{bmatrix} a_0 \\ a_1 \\ a_2 \\ \vdots \\ a_5 \end{bmatrix} = \begin{bmatrix} y_1 \\ y_2 \\ y_3 \\ \vdots \\ y_6 \end{bmatrix} \quad (17)$$

$$Z = \begin{bmatrix} 1 & 1 & 1 & - & - & 1 \\ \xi_{1,1} & \xi_{1,2} & \xi_{1,3} & - & - & \xi_{1,6} \\ \xi_{2,1} & \xi_{2,2} & \xi_{2,3} & - & - & \xi_{1,6} \\ \vdots & \vdots & \vdots & \vdots & \vdots & \vdots \end{bmatrix}$$

In the above equations, $Z$ can be calculated from $\xi$ at each sample point. Once the coefficients are estimated, the distribution of $Y$ is fully described by the Hermite polynomial expansion in equation (15). The statistical properties of the delay Y are calculated as described in Section 3.5.

## 4. Experimental Results

Our method has been tested extensively for several test cases and the results for some representative test cases are given below. BSIM3 model parameters of 0.18u technology are considered for all the logic gates in the examples. As a first example, we considered an Inverter and the uncertainty in $L_{eff}$ due to process variations. The mean of the PDF of $L_{eff}$ is taken as 0.18u and 20% variation as standard deviation (S.D). The $L_{eff}$ is truncated at 3σ points. The PDF of the 50% delay point is estimated using our proposed method and the spice based Monte Carlo (M.C) simulations. The comparison between our PCM method and the Monte Carlo method is presented in the Figure 3.
Table 2 presents the description of the examples we considered to test the PCM method and also gives the parameters in which we considered the variation. In all the test cases a variation of 10% to 20% is considered in the uncertain parameters. The mean values and standard deviation values estimated by our proposed method and Monte Carlo method for different examples are presented in Table 3. *Column 2* shows the mean values of the 50% delay PDF calculated from Monte Carlo method and *Column 3* shows mean values of the 50% delay PDF calculated from the our PCM method. *Column 4* shows the percentage error in mean values of our PCM method when compared with Monte Carlo method. *Column 5* and *column 6* shows the standard deviation values of 50% delay PDF calculated from Monte Carlo method and our PCM method respectively. *Column 7* shows the percentage error in standard deviation values of our PCM method when compared with Monte Carlo simulations method.

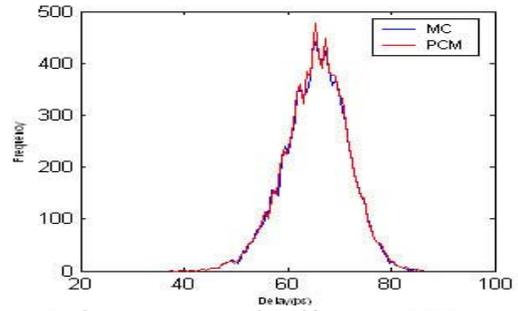

**Figure 3: Comparison of 50% delay PDF obtained with PCM and Monte Carlo for an inverter.**

**Table 2: Description of the examples considered to test our method**

| Example | Description | Variations Induced Parameters |
|---|---|---|
| 1 | Inverter | $L_{eff}$ |
| 2 | Inverter | $L_{eff}$, Tox, W |
| 3 | 2-Input NAND | $L_{eff}$, Tox |
| 4 | Two inverters in cascade | $L_{eff}$ |
| 5 | Full Adder | $L_{eff}$, Tox |
| 6 | NAND gate | Multiple Input Switching, $L_{eff}$ |

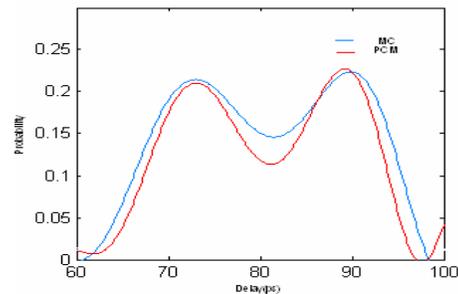

**Figure 4: Comparison of 50% delay PDF obtained with PCM and Monte Carlo for an inverter chain.**

In most of the cases, the gate delay PDF is close to Normal density function. But for larger circuits, the PDF is not symmetric. First, we considered an inverter chain of different sizes, and induced the variation of length with different standard deviation. The PDF is asymmetric and is shown in Figure 4. Second, we considered a full adder with variation in $T_{ox}$ and length. The PDF obtained by PCM is shown in Figure 5. Third degree polynomial expansion is used in all the cases. We tested our method for Multiple Input Switching case also. The arrival times at the inputs are considered as random variables. Variation in $L_{eff}$ due to process variations is considered. The PDF of 50% delay is shown in Figure 5. Second degree Hermite polynomial expansion is used for this example which resulted in relatively high error when compared to other test cases.

**Table 3: Comparison of mean and S.D of the PDF of 50% delay point obtained using PCM and Monte Carlo**

| Ex | Mean (ps) (M.C) | Mean (ps) (PCM) | % Error | S.D (M.C) | S.D (PCM) | % Error |
|---|---|---|---|---|---|---|
| 1 | 65.109 | 65.115 | 0.01 | 5.804 | 5.805 | 0.02 |
| 2 | 63.393 | 63.392 | -0.02 | 7.364 | 7.273 | -1.23 |
| 3 | 72.143 | 72.312 | 0.234 | 8.921 | 8.780 | -1.58 |
| 4 | 80.63 | 80.66 | 0.04 | 9.28 | 8.98 | -3.23 |
| 5 | 163.47 | 162.89 | -0.35 | 25.57 | 25.40 | -0.67 |
| 6 | 109.54 | 115.36 | 5.29 | 10.68 | 10.56 | 1.10 |

## 5. Conclusion

We proposed a novel method for modeling the uncertainty in the devices due to process variations and Multiple Input Switching. All the uncertain inputs are transformed into standard random variables and the output performance metric is expressed as a series expansion of orthogonal polynomials in terms of the standard random variables. We showed that Probabilistic Collocation method (PCM) accurately estimates the unknown polynomial coefficients which results in modeling the uncertainty accurately. We carried out extensive simulations of sample test cases. Comparison of the results from the proposed PCM technique against the Monte Carlo based spice simulations demonstrates an excellent match.

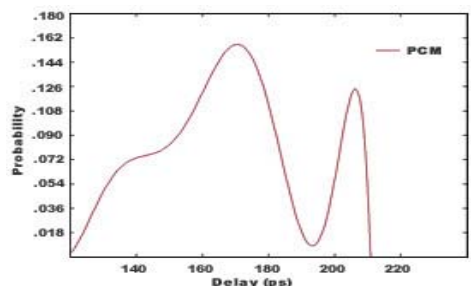

**Figure 5: 50% delay PDF obtained with PCM for Full Adder**

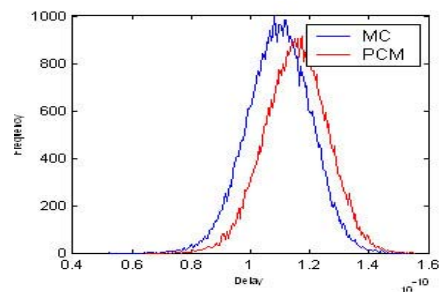

**Figure 6: Comparison of 50% delay PDF obtained with PCM and Monte Carlo for NAND gate considering Multiple Input Switching.**